\documentclass[%
 aip,
 amsmath,amssymb,
 reprint,%
]{revtex4-1}

\usepackage{graphicx}
\usepackage{dcolumn}
\usepackage{bm}
\usepackage{xcolor}  
\usepackage[utf8]{inputenc}
\usepackage[T1]{fontenc}
\usepackage{mathptmx}
\usepackage{etoolbox}

\makeatletter
\def\@email#1#2{%
 \endgroup
 \patchcmd{\titleblock@produce}
  {\frontmatter@RRAPformat}
  {\frontmatter@RRAPformat{\produce@RRAP{*#1\href{mailto:#2}{#2}}}\frontmatter@RRAPformat}
  {}{}
}%
\makeatother
\begin{document}

\preprint{AIP/123-QED}

\title[Dual-Frequency Comb in Fiber Fabry-Perot Resonator]{Dual-Frequency Comb in Fiber Fabry-Perot Resonator}
\author{Thomas Bunel}
 \email{thomas.bunel@univ-lille.fr; arnaud.mussot@univ-lille.fr}
\author{Debanuj Chatterjee}%
\affiliation{%
University of Lille, CNRS, UMR 8523-PhLAM Physique des Lasers, Atomes et Molécules, F-59000, Lille, France
}%

\author{Julien Lumeau}
\author{Antonin Moreau}
\affiliation{ 
Aix Marseille University, CNRS, Centrale Marseille, Institut Fresnel, Marseille, France
}%

\author{Matteo Conforti}
\author{Arnaud Mussot}
\affiliation{%
University of Lille, CNRS, UMR 8523-PhLAM Physique des Lasers, Atomes et Molécules, F-59000, Lille, France
}%
 
\date{\today}

\begin{abstract}
This paper presents a novel approach to dual-frequency comb generation utilizing a single fiber Fabry-Perot resonator, advancing the implementation of these sources in fiber-based systems. Dual-comb applications such as spectroscopy, ranging, and imaging, known for their high-resolution and rapid data acquisition capabilities, benefit significantly from the stability and coherence of optical frequency comb sources. Our method leverages the birefringent property of the resonator induced by the optical fiber to generate two orthogonally polarized optical frequency combs in a monolitic resonator. This approach allows for the generation of two different frequency combs with slightly different repetition rates, exhibiting excellent mutual coherence, making it highly relevant for dual-comb applications. The 40~nm bandwidth generated combs are induced by switching-waves in a normal dispersion fiber Fabry-Perot resonator. These comb types have the advantage of being easily generated by a pulse pumping scheme, which is employed in this study. Finally, the potential of the source is demonstrated by a proof-of-concept spectroscopy measurement.
\end{abstract}

\maketitle

%

\section{Introduction}
Optical frequency combs (OFC) offer broadband optical spectra comprised of many discrete frequencies that are equally spaced, corresponding to coherent train of pulses with a stable repetition rate~\cite{fortier_20_2019}. These precision light sources have become ubiquitous in various photonic technological applications~\cite{fortier_20_2019,torres-company_optical_2014}. Ultra-stable OFC-based spectroscopy, in particular, has seen significant growth in recent years due to the ability of these lasers to enhance the performance of existing spectrometers, such as virtual imaging phase array etalons, or Michelson-based Fourier transform interferometers~\cite{picque_frequency_2019,diddams_molecular_2007,mandon_fourier_2009}. Additionally, new spectroscopy techniques have emerged by using two OFC, significantly increasing analysis speed while maintaining high precision~\cite{coddington_dual-comb_2016,picque_frequency_2019}. This technique, named dual-comb spectroscopy, has attracted particular attention. It uses two frequency combs with slightly different repetition rates, which are interfered on a photodiode to generate a radio-frequency (RF) comb composed of distinguishable heterodyne beats between pairs of optical comb teeth. This RF comb is easily accessible with RF electronics and contains the relevant spectral information of the optical comb spectra~\cite{coddington_dual-comb_2016,picque_frequency_2019}. A key element for this technique is obtaining two frequency combs that are mutually coherent. Various technological systems have been developed to address this issue, including phase-locking of two mode-separated lasers~\cite{link_dual-comb_2015}, bidirectional lasers~\cite{ideguchi_kerr-lens_2016,Li:20,mehravar_real-time_2016}, arrays of electro-optic modulators (EOM) driven by a common laser~\cite{bancel_all-fiber_2023,parriaux_electro-optic_2020}, and dual micro-combs on a chip~\cite{bao_microresonator_2019,dutt_-chip_2018,suh_microresonator_2016,trocha_ultrafast_2018,Rebolledo_2023}. 
The latter have attracted significant attention over the last decade due to the high quality factors of the microresonators used, their compact design (with cavity lengths of hundreds of micrometers), and low energy consumption~\cite{sun_applications_2023}. Studies in this area have even shown that a dual-comb source can be achieved with a single resonator by utilizing two-way light circulation~\cite{suh_soliton_2018,zhang_soliton_2023}, two different transverse modes~\cite{lucas_spatial_2018}, or two polarization modes~\cite{xu_dual-microcomb_2021}. This simplifies the system by eliminating the need for two resonators and two pump lasers while also improving mutual coherence between the two OFC. Recently, fiber Fabry-Perot (FFP) resonators have emerged as a promising alternative for fiber systems~\cite{bunel_28_2024,nie_synthesized_2022,jia_photonic_2020,obrzud_temporal_2017,li_ultrashort_2023}. They exhibit high-quality factors, compact design, capability to generate OFC with line-to-line spacing in the GHz range, and simplified light coupling via FC/PC connectors (Ferrule Connector/Physical Contact). OFC generation has been demonstrated with these devices in anomalous~\cite{obrzud_temporal_2017,bunel_28_2024}, normal~\cite{bunel_broadband_2024,xiao_modeling_2023}, and even zero dispersion regimes~\cite{xiao_near-zero-dispersion_2023}. Broadband~\cite{bunel_28_2024} and highly stable~\cite{nie_synthesized_2022} frequency combs have also been demonstrated. However, while some studies have shown that it is possible to use the birefringence of fiber cavities to generate dual-combs with mode-locked lasers~\cite{ding_single-short-cavity_2023,zhao_polarization-multiplexed_2018}, to our knowledge, no dual-comb source exists with these passive cavities.

In this work, we report on the experimental demonstration of a dual-comb source based on switching-waves (SW) generation in a single normal dispersion FFP resonator. Using two electro-optically generated pulse trains derived from a single continuous wave (CW) laser, we simultaneously excite switching-wave combs in the two orthogonal polarization modes of the resonator. Thanks to the birefringence of the FFP cavity, the two polarization modes exhibit different free spectral ranges (FSR), thus allowing for the simultaneous generation of two OFC with different line spacings that are synchronized to the repetition rates of the respective driving pulse trains. Finally, we demonstrate the applied potential of the dual-comb source with a proof of concept for spectroscopy by measuring the transfer function of a programmable optical filter set as a notch filter.

\section{Experimental setup}\label{EXPSET}

\begin{figure}
\includegraphics{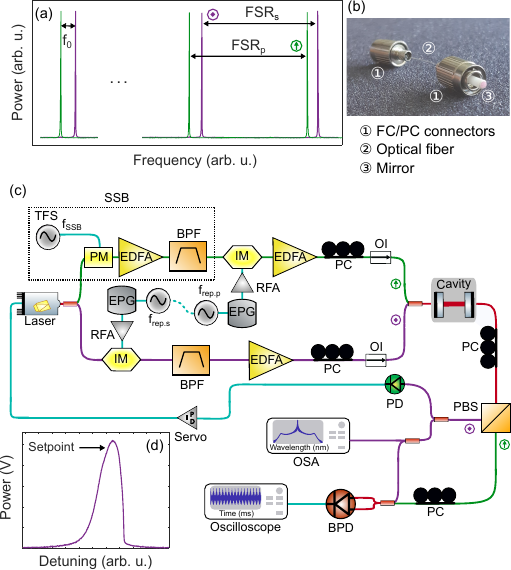}
\caption{Experimental setup. (a)~Representation of the resonance positions for both polarization mode; green line: polarization mode \textit{p}; purple line: polarization mode \textit{s}. (b)~Picture of the FFP cavity. (c)~Experimental setup schematic; green line: polarization mode \textit{p} beam; purple line: polarization mode \textit{s} beam. Both beams are perpendicularly polarized to each other. TFS: Tunable Frequency Synthesizer; EPG: Electrical Pulse Generator; RFA: Radio Frequencies Amplifier; IM: Intensity Modulator; PM: Phase Modulator; EDFA: Erbium Doped Fiber Amplifier; PC: Polarization Controller; OI: Optical Isolator; BPF: Bandpass filter; PD: Photodiode; PBS: Polarization Beam Splitter; PDH: Pound-Drever-Hall; SSB: single-side-band generator; OSA: Optical Spectrum Analyzer; BPD: Balanced Photodetector. (d)~Nonlinear transfer function of the polarization mode \textit{s} and locking system management.}
\label{setup}
\end{figure}

For this experimental study, a FFP cavity is constructed from an optical highly nonlinear fiber (HNLF Thorlabs-HN1550) with a length of $ L = 6.07 $~cm, a group velocity dispersion (GVD) of $\beta_2 = 1.3$~ps$^2$km$^{-1}$ at the pump wavelength (1550~nm), and a nonlinear coefficient of $\gamma = 10.8$~W$^{-1}$km$^{-1}$. Both fiber ends are mounted in FC/PC connectors, and Bragg mirrors are deposited at each end using a physical vapor deposition technique to achieve 99.86$\%$ reflectance over a 100~nm bandwidth \cite{zideluns_automated_2021} [see representation of the cavity in Fig.~\ref{setup}(b)]. This cavity design is particularly suitable for implementation in fiber systems thanks to the convenience offered by FC/PC connectors~\cite{obrzud_temporal_2017,bunel_observation_2023,bunel_28_2024}.
As depicted in Fig.~\ref{setup}(a), due to the residual birefringence of the resonator, it exhibits different FSR depending on the polarization mode. An FSR of $\text{FSR}_s \approx 1.696$~GHz is measured for polarization mode \textit{s}, while a slightly higher FSR is measured for polarization mode \textit{p}, $\text{FSR}_p = \text{FSR}_s+6$~kHz. Moreover, the resonances of both modes are offset by about $f_0 \approx 643 $~MHz at the pump wavelength of 1550~nm. For both modes, the finesse is the same: $\mathcal{F}=620$, resulting in a quality factor $ Q = 77 \times 10^6 $.
The normal-dispersion characteristic of the FFP resonator ($ \beta_2 > 0 $) favors the generation of switching-waves or dark solitons~\cite{xue_normal-dispersion_2016,fulop_high-order_2018}, unlike cavities with anomalous dispersion, which are more conducive to cavity soliton generation~\cite{godey_stability_2014}. Switching-waves have the advantage of being easily generated by a pulse pumping scheme~\cite{macnaughtan_temporal_2023,bunel_broadband_2024,xiao_modeling_2023}, and do not require specific and complex excitation protocols such as frequency kicking for cavity solitons~\cite{bunel_28_2024,kippenberg_dissipative_2018}. In the case of fiber cavities, which need a locking system to manage vibrations and thermal fluctuations, this is particularly advantageous as they can be generated using a simple PID (proportional-integral-derivative) servo-loop, unlike solitons, which often require a dual-arm stabilization system to pass through the chaotic state~\cite{bunel_28_2024,englebert_high_2023}. In our configuration, OFC can therefore be generated in each polarization mode, and stabilization is performed directly on the generated nonlinear signal of one mode~\cite{bunel_observation_2023}, using the edge of the resonance as an error signal [Fig~\ref{setup}(d)].

Thus, the FFP resonator is pumped by two orthogonally polarized pulse trains carved from a single ultra-narrow CW laser (0.1~kHz linewidth - \textit{BASIK E15} from \textit{NKT photonics}). The experimental setup is described in Fig~\ref{setup}(c). The laser beam is split into two paths, and a homemade single-sideband (SSB) generator, described in Ref.~\cite{bunel_28_2024}, is used to down-shift the frequency of light along the second path by $f_{SSB} = f_0 + 17 \times \text{FSR}_p = 29.484$~GHz~\cite{xu_dual-microcomb_2021}. Note that this value was chosen to enable the isolation of the sideband of the modulated beam. A frequency shift of $f_0$ would suffice if the technology permits. The pulsed pumps are generated using intensity modulators driven by electrical pulse generators to obtain gaussian pulses with a duration of $55$~ps full width half maximum. Each pump's repetition rate ($f_{\text{rep.s}}$ and $f_{\text{rep.p}}$ in Fig.~\ref{setup}(c)) is precisely adjusted using a frequency synthesizer to match the cavity roundtrip time within the Hz range for each polarization mode. Note that by generating switching-waves, we can determine if the repetition rate is perfectly synchronized by examining the symmetry of the generated comb~\cite{macnaughtan_temporal_2023,bunel_broadband_2024,xiao_modeling_2023}. Both frequency synthesizers are synchronized via a 10~MHz reference signal to prevent frequency drift between them. Erbium-doped amplifiers are used to increase the pump pulse peak power to 500~mW. In turn, polarization controllers ensure the polarization direction for each pump. At the output of the cavity, a polarization beam splitter (PBS) separates the two generated combs. The intensity of the \textit{s} polarized OFC is measured with a photodiode connected to the PID servo to adjust the laser frequency at a specific detuning, between the driving laser and the linear cavity resonance, as depicted in Fig~\ref{setup}(d). Note that the detuning of the \textit{p} polarized OFC will be exactly the same as the frequency offset $f_0$ is compensated by the SSB generator. An optical spectrum analyzer (OSA) can be used to monitor each comb. Finally, the orthogonally polarized combs are recombined using a 50/50 coupler, after tuning the polarization state of one OFC, to align it with the other one, generating an interferogram. This interferogram is then measured with a balanced photodetector and a 20~GHz bandpass oscilloscope.

\section{Optical frequency comb generation}

\begin{figure*}
\includegraphics{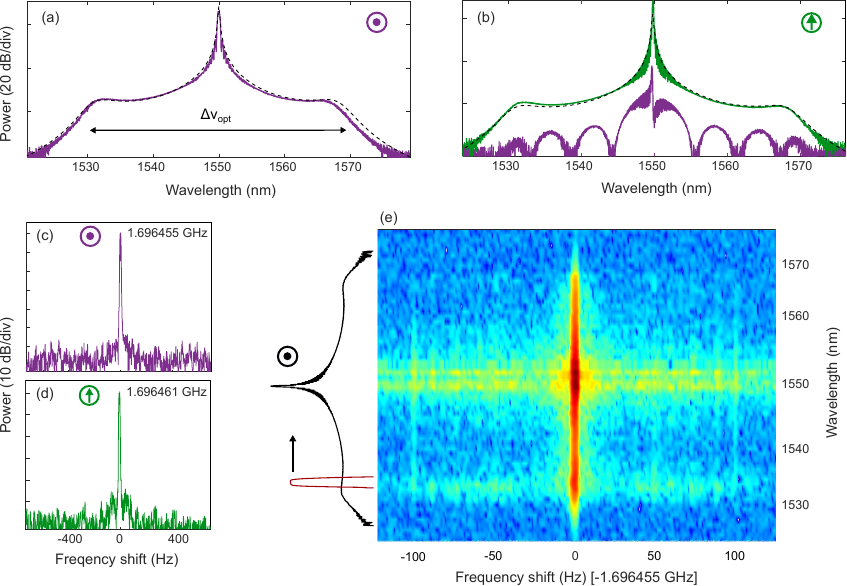}
\caption{Optical frequency combs in both polarization mode. (a)~Measurement on the polarization mode \textit{s}; purple line: measurement; black dashed line: numerics. (b)~Measurement on the polarization mode \textit{p}; green line: measurement; black dashed line: numerics; purple line: cross-talk between both polarization modes. (c) and (d)~Beatnote frequency of the generated combs, for polarization modes \textit{s} and \textit{p}, respectively; the center frequency of the beatnotes are noted at the top right, they differ by 6~kHz. (e)~Beatnote evolution along the generated comb; left panel: measurement principle, the beatnote is measured at the output of a bandpass filter which scans the comb from shorter to longer wavelength.}
\label{optics}
\end{figure*}

The generated OFC in both polarization modes are measured with OSA, and the results are depicted in Fig.~\ref{optics}(a) and (b). The setpoint of the PID servo is adjusted to reach a detuning of 0.023~rad for both pumps. Both combs exhibit symmetrical shoulders around the pump, a clear signature of switching-wave generation~\cite{macnaughtan_temporal_2023,bunel_broadband_2024,xiao_modeling_2023}. They span over more than 40~nm (\textit{i.e.}, 5~THz at 1550~nm) [purple and green lines in Fig.~\ref{optics}(a) and (b), respectively], and are in perfect agreement with simulations. These simulations are obtained by numerically solving an extended version of the Lugiato-Lefever equation adapted for a Fabry-Perot resonator~\cite{bunel_broadband_2024,cole_theory_2018,ziani_theory_2023}. 
Fig.~\ref{optics}(b) shows the cross-talk between the two polarization modes [purple line], measured by observing the transmitted signal in polarization mode \textit{p} while the pump in that mode is off, but on for polarization mode \textit{s}. We observe an extinction ratio between 20 and 40~dB, depending on the wavelength. This values fit well with PBS specifications for which a minimum extinction ratio of 20~dB is reported. Moreover, note that the interference pattern is common when using a PBS, suggesting that this component is the main cause of cross-talk rather than signal mixing within the cavity itself. With a single FFP cavity, we are able to generate two independent OFC with repetition rates difference of $\Delta f= FSR_p - FSR_s=f_{rep.p}-f_{rep.s}=6$~kHz, separated by an offset frequency of 643~MHz.

\begin{figure*}
\includegraphics{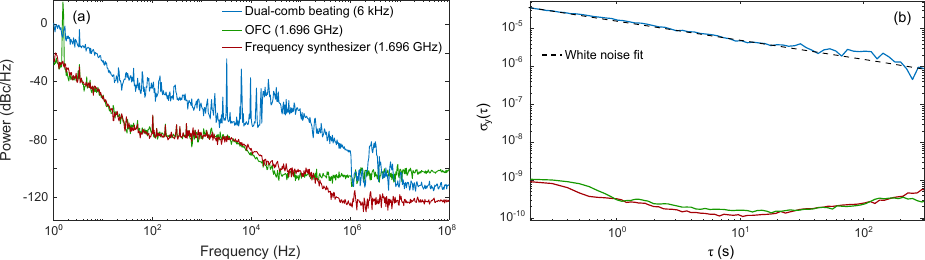}
\caption{Frequency stability. (a)~Phase noise spectra; (b)~Allan deviation. The legend in (a) applies to both boxes and includes the central frequency of the measurement in brackets.}
\label{stability}
\end{figure*}

The RF beatnotes of both OFC are shown in Fig.~\ref{optics}(c) and (d), exhibiting narrow linewidths that demonstrate the good stability of each generated OFC. As noted at the top right of the panels, the repetition rates are indeed different by 6~kHz.
The coherence of the combs are characterized by measuring the RF beatnotes of tens of comb teeth by filtering the generated signal of polarization mode \textit{s} with a tunable bandpass filter of 15~GHz bandwidth. The center frequency of this filter is tuned from one end of the comb spectrum to the other while continually measuring the RF beatnote. The results are shown in Fig.~\ref{optics}(e). The left part illustrates the measurement principle with the comb spectrum (in black) and the filter (in red). The colormap on the right shows the result. The beatnote consistently shows narrow linewidths, spanning less than 15~Hz, without degradation across the entire comb spectrum. This indicates that the corresponding repetition rate is extremely stable. Moreover, the stability of the comb teeth does not significantly degrade for large frequency shifts, since the beatnote linewidth does not vary, whether for teeth close to the pump or in the spectral wings, which is essential for using these signals in dual-comb applications. The same results are obtained with the generated OFC of polarization mode \textit{p}.

To gain further insight into the stability of the optical source, Fig.~\ref{stability} presents its performance in terms of phase noise [Fig.~\ref{stability}(a)] and long-term frequency stability, assessed through Allan deviation measurements [Fig.~\ref{stability}(b)]. The optical frequency combs are characterized by measuring the stability of one of the OFC beatnotes at $f_{rep}$ (1.696~GHz) when both OFC are generated simultaneously [green lines in Fig.~\ref{stability}]. The beatnote phase noise spectrum of the OFC polarized along \textit{s} mode, is measured using a photodiode and an electrical spectrum analyzer (ESA) [green lines in Fig.~\ref{stability}(a)]. It closely follows that of the frequency synthesizer driving the pump frequency [red line] (similar results are obtained with OFC polarized in mode \textit{p}). As demonstrated in previous studies on single SW-induced OFC generation in FFP resonators~\cite{xiao_modeling_2023,bunel_broadband_2024}, the OFC phase noise does not degrade compared to the pump, even when two combs are generated simultaneously. This behavior is also observed in long-term variations, as shown by the Allan deviation, measured with a frequency counter using a 100~ms gate time over a 20~minutes acquisition time. A local oscillator and a frequency mixer were also used to shift the signal frequency within the frequency counter bandwidth (300~MHz). The Allan deviation curves for the frequency synthesizer and the OFC [red and green lines in Fig.~\ref{stability}(b), respectively] are superimposed. This further supports the conclusion that the observed frequency instabilities primarily originate from the control electronics, as reported in references on EOM technologies~\cite{cai_design_2023,bancel_all-fiber_2023,parriaux_electro-optic_2020}. Moreover, the generated combs exhibit phase noise spectra and Allan deviation characteristics that are consistent with the state of the art for dual-comb sources in monolithic systems~\cite{bancel_all-fiber_2023,Li:20}, with -25~dBc/Hz at 1~Hz and -105~dBc/Hz at 1~MHz [green line in Fig.~\ref{stability}(a)], as well as a maximum stability plateau of $\sigma_y=1.5 \times 10^{-10}$ around 10~s [green line in Fig.~\ref{stability}(b)].

\section{Dual-comb source}

\begin{figure}
\includegraphics{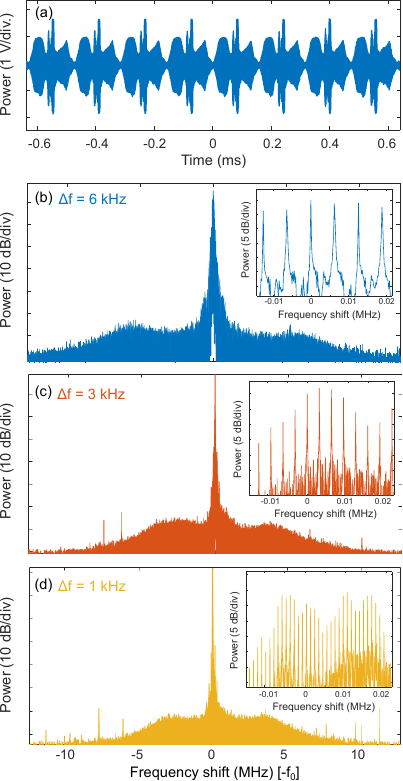}
\caption{Dual-comb source measurement in the RF domain. (a)~Multi-period interferogram between both combs in Fig.~\ref{optics}(a) and (b). (b)-(d)~Dual-comb spectra of varying line spacing with a zoom in the inset. The RF comb in (b) corresponds to the interferogram in (a)}
\label{rf}
\end{figure}

\begin{figure*}
\includegraphics{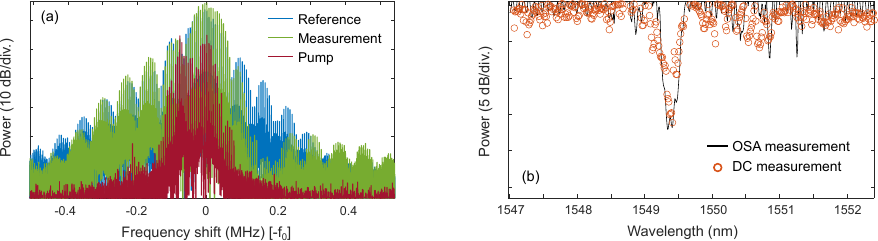}
\caption{Proof-of-concept spectroscopic measurement. (a)~Three RF combs; red lines: the pulsed pump, blue lines: the generated comb, green lines: the generated comb filtered by the \textit{Waveshaper}. (b)~Measurement of the \textit{Waveshaper} transfer function; black line: with the OSA, orange circles: with the dual-comb source.}
\label{spectro}
\end{figure*}

As explained in Section~\ref{EXPSET}, once the two combs have been generated, they are made to interfere by placing them on the same polarization axis and using a 50/50 coupler. This results in an interferogram in the time domain, which we measure using an oscilloscope with a sampling rate of $2$~GS.s$^{-1}$. Similar to other studies on dual-comb techniques~\cite{picque_frequency_2019,coddington_dual-comb_2016}, this interferogram consists of pulses with a repetition rate equal to $\Delta f=6$~kHz [Fig.~\ref{rf}(a)]. In contrast to dual-comb sources obtained with solitons, the interferogram in Fig.~\ref{rf}(a) does not exhibit sharp pulses. They contain two lobes on each side, probably due to the shape of the SW-induced OFC pulses, which differs from solitons in that they are much wider and rectangular in shape, with oscillations at the bottom of each front~\cite{macnaughtan_temporal_2023,xiao_modeling_2023,bunel_broadband_2024}. Therefore, the pulses are superimposed over a longer period, resulting in longer pulses in the interferogram. By performing a Fourier transform of this signal, we retrieve the shape of the optical spectrum in the RF domain, centered at a frequency of $f_0=643$~MHz [Fig.~\ref{rf}(b)], corresponding to the frequency shift between the two pumps~\cite{xu_dual-microcomb_2021}. Similar to the optical spectrum, the RF spectrum exhibits shoulders on both side of the pump component. The RF comb span can be theoretically calculated using the magnification factor $m=\frac{f_{rep}}{\Delta f}=\frac{\Delta \nu_{opt}}{\Delta \nu_{RF}}$, or the down conversion factor $\Delta \nu_{RF}=\frac{\Delta f}{f_{rep}}\Delta \nu_{opt}$, where $\nu_{opt}$ represents the optical frequencies and $\nu_{RF}$ the RF frequencies, leading to $\Delta \nu_{RF}=17$~MHz, with $\Delta \nu_{opt}=5$~THz and $f_{rep}=FSR_s=1.696$~GHz being the the optical frequency span and comb repetition rate, respectively, of the reference signal, which in our case is the comb of polarization mode \textit{s}. The measured RF comb corresponds to this value. The generated combs can be measured with a duration of tens of milliseconds using this dual-comb technique: 7~ms were necessary to record the spectrum in Fig.~\ref{rf}(b), compared to the few seconds required for traditional OSA measurements. In addition to increasing analysis speed, this technique significantly enhances spectral resolution, allowing the resolution of teeth separated by $FSR=1.696$~GHz, impossible to observe with a conventional OSA with a resolution of 12~GHz. 

The stability performance of this RF signal at the repetition rate difference $\Delta f$ (6~kHz) is also reported in Fig.~\ref{stability}. We observe a 20~dB increase in phase noise compared to the optical signal [blue line in Fig.~\ref{stability}(a)]. The peaks in the kHz range correspond to harmonics appearing at intervals of 6~kHz. The Allan deviation also exhibits a higher $\sigma_y$ [blue line in Fig.~\ref{stability}(b)], ranging from $10^{-6}$ to $3 \times 10^{-5}$, following a $1/ \sqrt{\tau}$ trend, characteristic of white frequency modulation noise~\cite{rubiola_companion_2023}.

The line spacing in the RF comb is directly related to the birefringence of the resonator, which can be tuned by applying mechanical stress to the fiber. Similar to methods employed for single-soliton generation~\cite{nie_synthesized_2022,jia_photonic_2020}, the birefringence of the FFP cavity can be adjusted by encapsulating it in a metal box and varying the pressure exerted by the lid using screws. This approach enables modification of the repetition rate difference between the two optical combs, thereby altering the line spacing of the generated RF comb. Two additional cases with line spacings of 3~kHz and 1~kHz are shown in Fig.~\ref{rf}(c) and (d), respectively. The insets in Fig.~\ref{rf}(b)-(d) illustrate the comb structures for the same window, showing the effective line spacing differences. As a result, the RF comb span varies with the line spacing, decreasing from 17~MHz when the repetition rate difference is 6~kHz [Fig.~\ref{rf}(b)], to 9~MHz at 3~kHz [Fig.~\ref{rf}(c)], and to 6~MHz at 1~kHz [Fig.~\ref{rf}(d)]. Furthermore, modifying the birefringence of the cavity also necessitates adjusting the frequency shift between the two pumps, which affects the center frequency of the RF spectrum $f_0$. Specifically, $f_0$ is 643~MHz in Fig.~\ref{rf}(b), 232~MHz in Fig.~\ref{rf}(c), and 328~MHz in Fig.~\ref{rf}(d).

Finally, to demonstrate the source's application potential, we use it to measure the spectral transmission profile generated by a programmable optical filter (\textit{Waveshaper}), set as a notch filter of 25~GHz FWHM bandpass. A symmetric configuration, with $\Delta f=6 kHz$, is used~\cite{coddington_dual-comb_2016}, which means that both frequency combs, after recombination, pass through the sample, here the programmable optical filter. Fig.~\ref{spectro}(a) shows the recorded RF combs before [blue lines] and after the programmable optical filter [green lines], which is placed after optical comb recombination. For comparison, the pump spectrum [red lines] is also measured using the dual-comb method to highlight the frequency broadening. First, we observe the sinc-like shape of the generated spectra [blue and green lines in Fig.~\ref{spectro}(a)], which is due to the square-like shape of the generated pulses at the cavity output, caused by the steep front involved in switching-waves~\cite{macnaughtan_temporal_2023,bunel_broadband_2024,xiao_modeling_2023}. Second, we notice the hole in the green spectrum, which is due to the programmable optical filter absorption. The original transfer function of the programmable optical filter can then be recovered in the optical domain by the operation $\nu_{opt}=\frac{c}{\nu_{RF}-f_0}\frac{f_{rep}}{\Delta f}+\frac{c}{\lambda_{pump}}$, where $\lambda_{pump}$ the pump wavelength, and $c$ the speed of light in vacuum. The absorption spectra in Fig.~\ref{spectro}(b) is obtained by normalizing the measured and reference electrical spectra, \textit{i.e.}, by subtracting the measured comb [in green in Fig.~\ref{spectro}(a)] from the reference comb [in blue] in the log-scale, divided by two as the configuration is symmetric~\cite{coddington_dual-comb_2016}. The results are then represented in the wavelength domain. The black curve is obtained using an OSA to measure the transfer function, while the orange circles are obtained using the dual-comb method. As can be seen, the dual-comb method works well and allows a measurement resolution equal to the FSR of the cavity, 1.696~GHz in our case, in 7~ms. It is worth noting that the measurement was taken close to the pump to benefit from a good signal-to-noise ratio. Although the entire optical spectrum could be used (including frequencies in the wings of the OFC), the low power in the comb teeth limits the accuracy of the measurements. A potential improvement could involve amplifying the signal at the cavity output~\cite{kuznetsov_ultra-broadband_2024,riemensberger_photonic_2022}, or designing an ultra-stable cavity, thermally and mechanically isolated, as opposed to the electrically stabilized setup used here.

\section{Summary and discussion}
In this study, we have presented an experimental demonstration of an orthogonally polarized dual-comb system in a monolitic FFP resonator. This type of cavity offers ease of integration into photonic systems through its FC/PC connectors, making it practical for generating OFC in the GHz range. By pumping two orthogonally polarized modes of a normal-dispersion FFP resonator with electro-optically generated pulse trains at different repetition rates, we were able to generate two coherent switching-wave-induced OFC with a line spacing difference of 6~kHz, each spanning 40~nm in bandwidth (approximately 5~THz at 1550~nm). The line spacing difference is directly related to the birefringence of the resonator, making it possible to tune this value by applying a mechanical stress to the fiber. This adjustment could either reduce acquisition time (by increasing the line spacing difference) or expand the accessible optical frequency range (by decreasing the line spacing difference).

By recombining the two generated OFC, we successfully retrieved the spectral shape of the optical frequency combs in the RF domain using a balanced photodetector and an oscilloscope. Moreover, we demonstrated the application potential of this source via proof-of-concept spectroscopic measurements by characterizing the transfer function of a programmable optical filter. The results achieved here were comparable to conventional techniques using a single source and OSA, while significantly improving acquisition speed, from several seconds to tens of milliseconds. While this is beyond the scope of the current study, amplification techniques for optical frequency combs may be a subject of future work to extend the range of applications. Therefore, the presented system is very relevant for dual-comb applications. It is worth noting that resolution and acquisition time can be tuned in these systems by changing the cavity length, or by using rational harmonic driving of the resonator to achieve optical combs with larger line spacing if desired~\cite{xu_harmonic_2020}. Additionally, FFP resonators have demonstrated their capability to generate cavity solitons~\cite{bunel_28_2024,jia_photonic_2020,obrzud_temporal_2017}. Dual-frequency comb sources based on this regime using these specific resonators could be an interesting area of research in the future. 

This work contributes to the development of new platforms for generating mutually coherent multi-frequency combs sources, with a view towards applications in fiber systems for spectroscopy, ranging, and imaging.

\begin{acknowledgments}
The present research was supported by the agence Nationale de la Recherche (Programme Investissements d'Avenir, ANR FARCO, ANR TRIPLE, LABEX CEMPI); Ministry of Higher Education and Research; European Regional Development Fund (WAVETECH),  the CNRS (IRP LAFONI); Hauts de France Council (GPEG project), and the university of Lille (LAI HOLISTIC).
\end{acknowledgments}

\section*{Conflict of interest}
The authors have no conflicts to disclose.

\section*{Data Availability Statement}
The data that support the findings of this study are available from the corresponding author upon reasonable request.

\nocite{*}
\section*{References}
\bibliography{references}

\end{document}